\documentclass[11pt, a4paper, onecolumn]{article}

\usepackage[utf8]{inputenc}
\usepackage{graphicx}
\usepackage{caption}
\usepackage{subcaption}
\usepackage{tabularx}
\usepackage{multirow}
\usepackage{array}
\usepackage{hyperref}
\usepackage{amsmath}
\usepackage{amssymb}
\usepackage{soul}
\usepackage[dvipsnames]{xcolor}
\usepackage{float}
\usepackage{mdframed}
\usepackage{authblk}
\usepackage[left=2.5cm,right=2.5cm, top=2cm, bottom=2cm]{geometry}

\newcommand{\QmostImpAsp}[0]{\textbf{Q1}}
\newcommand{\QaddEffPoly}[0]{\textbf{Q2}}
\newcommand{\QavgDiaPoss}[0]{\textbf{Q3}}
\newcommand{\QimpBouCond}[0]{\textbf{Q4}}
\newcommand{\QfurExpReq }[0]{\textbf{Q5}}

\newcommand{\diff}[0]{\mathrm{d}}
\newcommand{\difft}[1]{\frac{\diff #1}{\diff t}}
\newcommand{\difftinline}[1]{\diff #1/\diff t}
\newcommand{\diffxinline}[1]{\diff #1/\diff x}
\newcommand{\qconv}[0]{\dot{q}_{\mathrm{conv}}}
\newcommand{\qrad}[0]{\dot{q}_{\mathrm{rad}}}

\newcommand{\inv}{^{-1}}
\newcommand{\kelvin}{K}
\newcommand{\Pa}{Pa}

\newcommand{\meter}{m}
\newcommand{\mum}{\mu\meter}
\newcommand{\kg}{kg}
\newcommand{\squaremeter}{m^2}

\newcommand{\second}{s}
\newcommand{\persecond}{\second\inv}
\newcommand{\permeter}{\meter\inv}
\newcommand{\persquaremeter}{\meter^{-2}}
\newcommand{\percubicmeter}{\meter^{-3}}
\newcommand{\flux}{\persecond\persquaremeter}

\newcommand{\centi}{c}
\newcommand{\milli}{m}

\newcommand{\unit}[1]{\ensuremath{\mathrm{#1}}}
\newcommand{\num}[1]{\ensuremath{\mathrm{#1}}}
\newcommand{\qty}[2]{\num{#1}\ensuremath{\,}\unit{#2}}

\begin{document}

\title{A comparison of mechanistic models for the combustion of iron microparticles and their application to polydisperse iron-air suspensions}
\author[*1]{J. Mich}
\author[1]{D. Braig}
\author[23]{T. Gustmann}
\author[1]{C. Hasse}
\author[1]{A. Scholtissek}
\affil[*]{\small Corresponding author: mich@stfs.tu-darmstadt.de}
\affil[1]{\small Institute for Simulation of Reactive Thermo-Fluid Systems, TU Darmstadt, \newline Otto-Berndt-Stra{\ss}e 2, 64287 Darmstadt, Germany}
\affil[2]{ \small IFW Dresden, Institute for Complex Materials, 01069 Dresden, Germany}
\affil[3]{\small Present affiliation: OSCAR-PLT-GmbH, Klipphausen 01665, Germany}
\date{June 2023}
\maketitle

\begin{abstract}

   Metals can serve as carbon-free energy carriers, e.~g.~in innovative metal-metal oxide cycles as proposed by Bergthorson~(Prog. Energy Combust. Sci., 2018). For this purpose, iron powder is a suitable candidate since it can be oxidized with air, exhibits a high energy density, is non-toxic and abundant.
   Nevertheless, the combustion of iron powder in air is challenging especially with respect to flame stabilization which depends on the particle size distribution and the morphology of the iron microparticles among other factors.
   Models for the prediction of reaction front speed in iron-air suspensions can contribute to overcoming this challenge. To this end, three different models for iron particle oxidation are integrated into a laminar flame solver for simulating such reaction fronts.
   The scientific objective of this work is to elucidate the influence of polydispersity of the iron particles on the reaction front speed, which is still not satisfactorily understood. In a systematic approach, cases with successively increasing complexity are considered: The investigation starts with single particle combustion and then proceeds with iron-air suspensions prescribing binary particle size distributions (PSDs), parametrized generic PSDs, and an exemplary PSD measured for a real iron powder sample.
   The simulations show that, dependent on the kinetic particle model, the particles' thermal inertia, and the PSD, particles undergo thermochemical conversion in a sequential manner according to their size and every particle fraction exhibits an individual combustion environment (surrounding gas phase temperature and oxygen concentration).
   The local environment can be leaner or richer than the overall iron-to-air ratio would suggest and can be very different from single particle experiments. The contribution of individual particle fractions to the overall reaction front speed depends on its ranking within the PSD. The study further demonstrates, that although the three particle models show good agreement for single particle combustion, they lead to very different reaction front speeds. This is due to the different ignition behaviour predicted by the particle models, which is shown to strongly influence the reaction front characteristics.
   Overall, the present work illustrates the complex relationship between characteristics of single particle ignition and combustion, polydispersity, and properties of reaction fronts in iron-air suspensions. 

\end{abstract}
{\footnotesize \textbf{Keywords:} metal fuels, polydispersity, solid fuel combustion, Euler-Lagrange modeling, reaction front speed}

\section{Introduction}
\label{sec:introduction}

The transformation of the energy sector to a sustainable energy economy is one of the primary challenges of our time~\cite{UN17Goals}. While clean, renewable power generation from solar and wind are commercially available and cost-competitive, their volatility and geographic availability require large scale energy storage solutions for energy trade and transportation, as well as seasonal on-demand energy supply. In this context, iron is discussed as a chemical energy carrier which could help overcome such challenges~\cite{Bergthorson2015, Julien2017}, most notably due to the possibility of retrofitting existing coal infrastructure for carbon-free operation with iron~\cite{Debiagi2022}.
Besides other processes to utilize the energy stored in iron~\cite{Bergthorson2017}, the combustion of micron-sized iron particles in air is promising due to the release of heat at high temperatures which can be directly used in power generation and industrial processes~\cite{Bergthorson2018}. However, the combustion of iron and other metal fuels greatly differs from the conventional hydrocarbon fuels~\cite{Bergthorson2015, Goroshin2022}. Although a lot of progress has been made recently with respect to the understanding of iron combustion, our understanding of the subject is still far from being complete or comprehensive and many open research questions remain~\cite{Goroshin2022}.

An appropriate modeling of single particle combustion is widely considered to be the basis for the simulation of iron flames. It has been shown in the literature that iron particles exhibit a thermal runaway in oxidizing atmosphere when heated beyond a threshold temperature (i.~e.~the ignition temperature), such that the particle temperature overshoots and  vastly exceeds the surrounding gas temperature~\cite{Soo2015, Soo2017}. At low temperatures, the reaction of iron with oxygen is limited by chemical kinetics balancing heat dissipation into the surroundings~\cite{ Soo2015, Mi2022}. With increasing temperature, the chemical kinetics accelerate until, at the ignition temperature, the heat release from the chemical reaction exceeds the heat dissipation to the gas environment subsequently leading to a rapid increase of the particle temperature~\cite{Soo2015}. During this thermal runaway phase, very high reaction rates are reached, which become limited by the external diffusion of oxygen through the particle boundary layer~\cite{Mi2022, Ning2021, Thijs2022a}. In this context, Goroshin~\cite{Goroshin2022} introduced the notion of considering individual particles as "micro-reactors".

Over the last few years, several approaches have been proposed to describe this single particle combustion behaviour with models of different complexity. Qualitatively, different expressions are used with respect to the initiation of the thermal runaway:

Goroshin~\cite{Goroshin1996} developed an analytical particle model, which qualitatively captures the preheating, thermal runaway and diffusive burnout. In Goroshin's model, the particle is considered chemically inert at first and the thermal runaway is initiated as soon as the particle reaches the previously specified ignition temperature. Afterwards, the particle oxidation proceeds at a fixed reaction rate. The model was later extended by Goroshin and co-workers~\cite{Goroshin2000, Palecka2018} to also consider binary fuel mixtures and was used to investigate discrete effects in two-dimensional~\cite{Goroshin2011a} and three-dimensional~\cite{Tang2012, Lam2020} flames.

Soo~et~al.~\cite{Soo2015} proposed a model which considers the rate limitation by both heterogeneous reaction kinetics and mass transfer. It uses an Arrhenius-type, first order surface reaction to model the kinetics. Thereby, the reaction rate is linearly dependent on the oxygen concentration on the particle surface. As a result, the particle ignition temperature is inversely correlated to the particle diameter~\cite{Mi2022}. The boundary layer diffusion is modeled according to a semi-empirical Sherwood relationship. Several extensions have been added to the particle model in the last years to include, for instance, Stefan flow effects or the formation of nanoparticles~\cite{Thijs2022a}.

Recently, Mi~et~al.~\cite{Mi2022} proposed a model derived from physical insights into the solid phase iron oxidation mechanism governing the chemical kinetics. In their approach, they assume that the kinetically limited combustion phase depends on the transport of iron cations from the particle core to the phase interfaces, where the reactions take place. The model is based on Wagner's theory~\cite{Wagner1933}, which accounts for the growth of oxide layers determining the solid state diffusion resistance during particle oxidation~\cite{Mi2022}. During particle conversion, the diffusion resistance increases proportionally to the thickness of the forming oxide layers. Thereby, the reaction rate follows a semi-empirical parabolic rate law, analogous to experimental observations from the corrosion of planar iron samples~\cite{Paidassi1958, Atkinson1985}. According to the model by Mi~et~al.~\cite{Mi2022} the ignition temperature is independent from the diameter for large particles.

While the laminar burning velocity is an important reference quantity for the combustion of gaseous fuels, only few experimental measurements of reaction front speeds in iron-air suspensions are available up to now. Some authors measured the reaction front speeds for iron-air suspensions in gravity reduced environments~\cite{Tang2009a, Goroshin2011b, Palecka2019}, however, many of these experiments were designed to study discrete effects in iron dust flames and did not focus on the continuous flame propagation regime. Therefore, instead of using the reaction front speed as the key observable, the aforementioned particle models are usually calibrated and validated using experimental data from single particle combustion experiments. To this end, Ning~et~al.~\cite{Ning2021, Ning2022} provided measurements of the particle temperature during combustion for varying particle sizes and oxygen contents in the sourrounding gas phase. Recently, Li~et~al.~\cite{Li2022} reported similar temperature profiles for porous particles. The reported datasets primarily cover the high temperature combustion phase (which is largely diffusion-limited) since the pyrometry measurement techniques utilized in the experiments are usually limited to temperatures above~\qty{1500}{\kelvin}~\cite{Ning2022}. 

Besides calibrating models with such single particle experiments, Goroshin~et~al.~\cite{Goroshin2022} emphasized that also the kinetically limited ignition behavior of particles needs to be considered when modeling iron dust flames. According to their work, the ignition temperature is not an external boundary condition, but part of the solution since it depends on the reaction front structure, especially in polydisperse metal powders. Therefore, they pointed out the need for detailed numerical analyses of reaction front propagation in realistic, polydisperse metal powder suspensions.

A first numerical investigation of effects induced by polydispersity was recently presented by Ravi~et~al.~\cite{Ravi2022}. The authors showed that the maximum reaction front speeds in binary powder suspensions can be reached at different equivalence ratios depending on the diameter ratio. They explained this effect by the overlapping or separation of flame fronts, which is consistent with earlier analytical work by Goroshin~et~al.~\cite{Goroshin2000, Palecka2018}. Their results underline the importance of polydispersity, however, the interplay of particle kinetics, thermal runaway, and polydispersity inquired by Goroshin~et~al.~\cite{Goroshin2022} still requires further investigation and represents the motivation for this study.

In our work, we employ the state of the art particle models coupled with a one-dimensional gas phase model to investigate the influence of the chemical kinetics on the reaction front speed. Particular focus is put on the linkage between the ignition behavior predicted by the individual particle models and the effects from polydispersity. With this background, it is the objective of this work to answer the following scientific questions:

\begin{itemize}
    \item[\QmostImpAsp{}] Which characteristics of single particle combustion are most influential for the reaction front speed?
    \item[\QaddEffPoly{}] How does polydispersity influence the reaction front speed?
    \item[\QavgDiaPoss{}] Is it possible to approximate the reaction front speed for a polydisperse iron-air suspension with a monodisperse equivalent?
    \item[\QimpBouCond{}] Which additional parameters need to be accurately determined in experiments and modelled in simulations to recover reaction front speeds?
    \item[\QfurExpReq{}] Which further experimental reference cases are required?
\end{itemize}

The paper is structured as follows: Sec.~\ref{sec:methods} describes the numerical modeling approach followed by a verification of the single particle models in Sec.~\ref{sec:verification}. In Sec.~\ref{sec:resultsAndDiscussion}, the influence of polydispersity on reaction front speeds in iron particle suspensions is analyzed systematically by means of numerical simulations considering generic and realistic particle size distributions. Results from the different models are compared to each other, revealing the influence of the kinetic modeling strategies on the reaction front speed. The paper ends with a summary and conclusion in Sec.~\ref{sec:conclusion}.

\section{Numerical methods}
\label{sec:methods}

\begin{figure}[htb]
    \centering
    \includegraphics[scale=1]{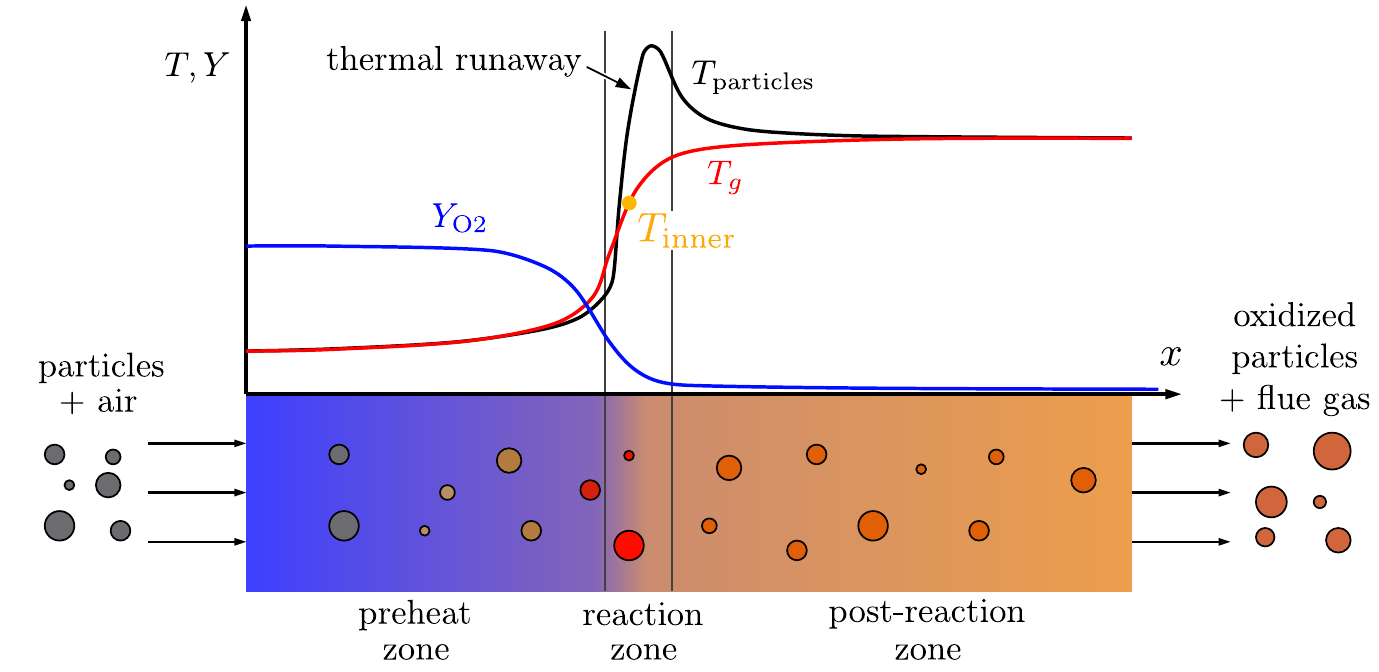}
    \captionsetup{justification=centering}
    \caption{Schematic of the one-dimensional numerical setup. Suspended iron particles in the oxidizer enter the domain from the left. The iron particles are oxidized in the reaction zone. The flue gas stream carrying oxidized particles leaves the domain at the outlet to the right. The colors in the bottom part represent the temperature evolution of gas and particles from cold (blue/gray) to hot (red).}
    \label{fig:setup}
\end{figure}

In order to determine the reaction front speed for various conditions in iron-air suspensions, canonical 1D freely-propagating iron-air flames are computed with an Euler-Lagrange multiphase modeling approach. The numerical configuration is visualized in Fig.~\ref{fig:setup}. An oxidizer (air) stream carrying disperse iron particles of different sizes enters the domain at the inlet and a hot flue gas stream carrying oxidized particles leaves the domain at the outlet. For the gas phase, temperature and species mass fractions are obtained as a result from the numerical simulation. The system is considered isobaric with a constant pressure of $p=\qty{101325}{\Pa}$. Furthermore, an inner boundary condition is applied inside the reaction zone, which fixes the temperature: $T(x=\qty{0.01}{\meter}) = T_{\mathrm{inner}} = \qty{1000}{K}$, see Fig.~\ref{fig:setup}. With that, the position of the reaction front in the gas phase (i.~e. large temperature gradient, red line in Fig~\ref{fig:setup}) is stabilized inside the computational domain. Similar approaches are used by previous authors in 1D numerical setups~\cite{Filho2018, Hazenberg2021}. At the inner boundary condition with known temperature, a Dirichlet boundary condition is obtained for the mass flux, which is an eigenvalue of the numerical problem.
In the high temperature reaction zone, the particles ignite and undergo a thermal runaway \cite{Soo2015, Soo2017}, as indicated with the black line in Fig.~\ref{fig:setup}, releasing their chemical energy as heat. In the following, the submodels for gas and disperse solid phase are described in detail.

\subsection{Gas phase}
\label{sec:gasPhase}

The steady-state Eulerian gas phase equations for mass flux~$\dot{m}_g$, temperature~$T_g$, and species mass fraction~$Y_i$ read \cite{Book-Poinsot-NumericalComb, Book-Chung-CombustionPhysics}:
\begin{align}
    \frac{\diff \, \dot{m}_g}{\diff x} & = \dot{m}_t, \label{eq:gas:mflux}\\
     c_{p,g} \frac{\diff \, \dot{m}_g T_g}{\diff x}  & = \sum_{i=1}^N\omega_i h_i + \frac{\diff}{\diff x} \,  \left(\lambda_g \frac{\diff T_g}{\diff x}\right) + \dot{h}_t - \varrho_g \sum_{i=1}^{N} \, c_{p,i} Y_i V_i \, \frac{\diff T_g}{\diff x}, \label{eq:gas:temperature} \\
     \frac{\diff \, \dot{m}_g Y_i}{\diff x} & =\omega_i -  \frac{\diff}{\diff x} \, \left( \varrho_g Y_i V_i\right) + \dot{Y}_{i.t} , \; i=1,...,N. \label{eq:gas:species}
\end{align}
Body forces, heat losses, and shear stresses are neglected. Quantities which describe the properties of the gas mixture, are denoted by the subscript~$g$, while quantities with subscript~$i$ refer to species-specific properties. The mass flux, $\dot{m}_g$, is related to the gas phase velocity~$u_g$ via the density $u_g = \dot{m}_g / \varrho_g$.
The gas phase density~$\varrho_g$ is obtained from the ideal gas law. The thermodynamic properties, i.~e. the specific enthalpy~$h_i$ and heat capacity~$c_{p,g}$ as well as the transport properties species diffusion velocity~$V_i$ and reaction rate~$\omega_i$ are calculated with Cantera~\cite{CanteraV251} using the GRI~3.0 mechanism~\cite{smith_1999} and a mixture-averaged diffusion model~\cite{Book-Hirschfelder1964}.

In this paper, air ($21\,\%$~$\mathrm{O2}$, $78\,\%$~$\mathrm{N2}$, $1\,\%$~$\mathrm{Ar}$) with an initial temperature $T_0=\qty{300}{\kelvin}$ is used as oxidizer for all calculations. These values are fixed as Dirichlet boundary conditions at the inlet. A zero-gradient boundary condition $\diffxinline{T} = \diffxinline{Y_i}= \num{0}$ is used at the outlet.
The gas phase is coupled to the iron particles by exchange of heat and mass and the particles are solved separately in an operator splitting approach. The mass transfer is described by~$\dot{Y}_{i,t}$ for every species~$i$,
the total mass transfer of all species is given by~$\dot{m}_t$. Finally, $\dot{h}_t$ is the heat transfer between gas and solid phase. The transfer terms are per gas volume and time, as specified in Sec~\ref{sec:numericalSolution}.

\subsection{Disperse solid phase}
\label{sec:solidPhase}

The particles are tracked in a Lagrangian manner in the one-dimensional domain and are coupled to the gas phase by means of a particle-in-cell (PIC) method. To track a particle through the domain, the Lagrangian particle equations are integrated over time with the boundary conditions given by the gas phase state at the particle's current position. In polydisperse simulations, every particle fraction has its own set of properties, i.~e. different initial masses and diameters.
Further details about the numerical solution algorithm are provided in Sec.~\ref{sec:numericalSolution}. For the formulation of the equations, we consider the particle boundary layer as a continuous phase. Recent work by Senyurt~et~al.~\cite{Senyurt2022} suggests that this assumption may be a simplification for metal combustion and that the Knudsen regime can influence particle ignition and peak temperatures. Our work does not include these effects such that future investigations on the interplay between polydispersity and boundary layers in the transition regime are required. The continuous equations for the particle's sensible enthalpy~$H_p$ \cite{Book-Chung-CombustionPhysics} and velocity~$u_p$ are:
\begin{align}
    \difft{H_p} & = \qconv + \qrad + \difft{m_p} h_{\mathrm{O2},s},
    \label{eq:particles:enthalpy} \\
    \qconv & = A_p \frac{\mathrm{Nu} \lambda_f}{d_p} \, (T_p - T_g) \frac{\mathrm{ln}(1+B_t)}{B_t},\label{eq:qconv} \\
    \qrad & = \epsilon \sigma \, A_p \, (T_p^4 - T_g^4),\label{eq:qrad}\\
    B_t & = \frac{Y_{\mathrm{O2},s}-Y_{\mathrm{O2},g}}{1 - Y_{\mathrm{O2},s}}, \\
    \difft{u_p} & = -\frac{1}{2}C_d   \, \varrho_f \frac{\pi d_p^2}{4}  \, |u_g - u_p|(u_g - u_p), 
    \label{eq:particles:velocity} \\
    C_d & = \frac{24}{\mathrm{Re}_p} \, \left(1  +0.15 \mathrm{Re}_p^{0.687}\right),
    \label{eq:particles:DragCoefficient}
\end{align}
where the expression for the drag coefficient, $C_d$, according to Schiller and Naumann~\cite{Schiller1933} is used. The particle quantities are denoted by the subscript~$p$ and the subscript~$f$ refers to properties of the boundary layer surrounding the particle. The latter are evaluated with Cantera~\cite{CanteraV251} using the same species mixture and pressure as the bulk gas at an averaged boundary layer temperature determined by the one-third rule, $T_f = 1/3\,T_g + 2/3\, T_p$. $Y_{\mathrm{O2},s}$ is the oxygen concentration on the particle surface, which is determined from the oxygen consumption rate. Also $h_{\mathrm{O2},s}$, the enthalpy of the consumed oxygen, is evaluated with Cantera at the particle temperature. $A_p=\pi d_p^2$ is the particle outer surface with the particle diameter calculated from its mass and density: $d_p =(6 m_p / \varrho_p /\pi)^{1/3}$. The particle Reynolds number~$\mathrm{Re}_p$ is given by $\mathrm{Re}_p = |u_g - u_p| d_p / \nu_f$. Heat transfer between particles and gas phase involves a convective term $\qconv$, described by a Nusselt relationship with a constant Nusselt number $\mathrm{Nu} = 2$, and a radiative term $\qrad$, which depends on the emissivity~$\epsilon$ and the Boltzmann constant~$\sigma$. In the convective heat exchange (Eq.~\ref{eq:qconv}), the influence of the Stefan flow is included~\cite{Book-Chung-CombustionPhysics}. 

To calculate the temperature from the particle enthalpy, tabulated thermodata from the NIST Chemistry WebBook~\cite{NIST_Chase1998} is used. The table is extended to also involve the fusion enthalpy of the different oxidation states. It is worth mentioning that $\mathrm{FeO}$ denotes a crystalline solid structure which does not have a corresponding liquid state. However, as an approximation, during melting FeO is assumed to transition to liquid iron oxide with a molar ratio $1:1$ between $\mathrm{Fe}$ and $\mathrm{O}$ for which also thermodynamic data is available.

The state variables describing the particle and the evolution of the particle mass $\mathrm{d}m_p/\mathrm{d}t$ are specific to the individual particle models which are described next.

\subsubsection{Fixed Reaction Rate (FIRR) Model}
\label{sec:FIRR}

The first particle model is based on the analytical model by Goroshin~et~al.~\cite{Goroshin1996}, which assumes a fixed reaction rate (FIRR) after the particle temperature exceeds the ignition temperature. For the model, the ignition temperature, $T_{\rm{ign}}$, and the burnout time, $\tau_c$, are calibration parameters. The reaction is stopped either by reaching full conversion or by the depletion of all oxygen in the gas phase surrounding the particle. The original model is translated into a Lagrangian formulation here. The equation for the rate of mass change then reads:
\begin{align}
    \difft{m_p} & = 
    \begin{cases}
        m_{p,\mathrm{Fe},0} \, s / \tau_c & \textbf{if} \quad\; T_p > T_{\rm{ign},} \; Y_{\mathrm{O2},g} > 0, \\
        0 & \textbf{else}.
    \end{cases}
    \label{eq:FIRR:mass}
\end{align}
In the above equation, $s$ is the stoichiometric ratio of the oxidation of iron to iron oxide: $\mathrm{Fe + s O \to FeO_s}$ \cite{Hazenberg2021}. The unit of~$s$ is \unit{\kg} oxygen per \unit{\kg} iron and $m_{p,\mathrm{Fe},0}$ is the initial mass of iron in the particle. Notably, the FIRR model does not incorporate a submodel for the mass transfer through the particle boundary layer, such that a diffusion limitation cannot be accounted for. 

\subsubsection{First Order Surface Kinetics (FOSK) Model}
\label{sec:FOSK}

The second particle model is based on the model by Soo et al.~\cite{Soo2015, Soo2017}. It combines a First Order Surface Kinetics (FOSK) model with a model for the oxygen diffusion through the particle boundary layer, given by a Sherwood relationship. Further model extensions have been developed recently by other authors~\cite{Thijs2022a, Hazenberg2021}. The stoichiometric factor $s$ is set equal to the FIRR model. The governing equations for the model read:
\begin{align}
    \difft{m_p} & = \dot{m}_{\mathrm{O2,kin}} = \dot{m}_{\mathrm{O2,diff}}, \label{eq:FOSK:mass}\\
    \dot{m}_{\mathrm{O2,kin}} & = Y_{\mathrm{O2},s} A_p k_{\infty}\, \mathrm{exp} \left( -\frac{T_a}{T_p}\right), \label{eq:FOSK:kinetic}\\
    \dot{m}_{\mathrm{O2,diff}} & = \varrho_f \, A_p \frac{\mathrm{Sh}\, D_{\mathrm{O2},f}}{d_p} \mathrm{ln}\left(1+B_m\right), \label{eq:FOSK:diffusion}\\
    B_m &= B_t = \frac{Y_{\mathrm{O2},s}-Y_{\mathrm{O2},g}}{1 - Y_{\mathrm{O2},s}}.
\end{align}

If the particle conversion rate is kinetically limited, the Arrhenius expression in Eq.~(\ref{eq:FOSK:kinetic}) determines the particle conversion rate. In this case, the oxygen concentration on the particle surface $Y_{\mathrm{O2},s}$ approaches the far field concentration and $B_m$ tends to zero.
In Eq.~(\ref{eq:FOSK:kinetic}), $T_a$ is the activation temperature and $k_{\infty}$ the pre-exponential factor~\cite{Soo2015}. In case of diffusion limitation, $Y_{\mathrm{O2},s}$ approaches zero and the boundary layer diffusion determines the particle conversion rate, modeled with a Sherwood relationship according to Eq.~(\ref{eq:FOSK:diffusion})~\cite{Soo2015}, where $D_{\mathrm{O2},f}$ is the diffusion coefficient of oxygen in the boundary layer. The Sherwood relationship involves the film layer density~$\rho_f$. Early works on iron particles neglected the density gradient in the boundary layer, approximating it by the bulk gas density~$\rho_g$ \cite{Soo2015, Mi2022, Hazenberg2021}. Recently, Thijs~et~al.~\cite{Thijs2022a} showed that due to the high density gradients induced by the temperature difference between particles and bulk gas, an averaged film density leads to more accurate predictions, which is also consistent with the original theory by Ranz and Marshall~\cite{Ranz1952}.

\subsubsection{Solid Oxide Layer Diffusion (SOLD) Model}
\label{sec:SOLD}

Thirdly, the particle model developed by Mi~et~al.~\cite{Mi2022} is extended to be used in the present work. In the kinetically limited regime, the particle conversion rate depends on the solid oxide layer diffusion (SOLD) of iron cations as described in Sec.~\ref{sec:introduction}. Opposite to the first order surface kinetics (FOSK) model, the oxide layer diffusion represents a different physical mechanism governing the iron particle ignition (the scope of the work by Mi~et~al~\cite{Mi2022}). Subsequent to ignition and thermal runaway, particles melt into a liquid droplet, the oxide shell structure disappears, and the internal kinetics become governed by other processes. Recently, it was suggested by Thijs et al.~\cite{Thijs2023} that the surface chemisorption and internal transport could become the rate limiting step at high temperatures. So far, a particle model incorporating these effects in a comprehensive manner is not yet available and therefore not further discussed here. The models in this work assume that the reaction rate at high temperatures is primarily controlled by external diffusion of oxygen through the particle boundary layer. The exponential temperature dependency of internal kinetics leads to kinetic rates exceeding the boundary layer diffusion above roughly~\qty{1300}{\kelvin} to~\qty{1400}{\kelvin} such that the latter becomes rate limiting. We can therefore extrapolate the solid state kinetics also to the liquid state, without notably affecting the particle burnout times.

In this work, contrary to the original approach by Mi~et~al.~\cite{Mi2022}, the formation of $\mathrm{Fe_3O_4}$ is excluded in order to achieve comparability with the models FIRR and FOSK. To describe the complete burnout of the particles including layer thicknesses up to the particle radius, a curvature correction is introduced based on the theory by Fromhold~et~al.~\cite{Fromhold1988}. A detailed description and derivation of the curvature correction is given in the \hyperlink{appendix}{Appendix}. With these modifications, the equations for the SOLD model read:

\begin{align}
    \difft{m_{\mathrm{FeO}}} & = 
        \frac{4 \pi \, r_{\mathrm{FeO}} r_{\mathrm{Fe}}}{r_{\mathrm{FeO}} - r_{\mathrm{Fe}}} \,
        \varrho_{\mathrm{FeO}} \, k_{\infty, \mathrm{FeO}} \, \mathrm{exp}\left(-\frac{T_{a,\mathrm{FeO}}}{T_p}\right)
        \label{eq:SOLD:FeO_curv}\\
    \difft{m_{\mathrm{Fe}}} & = 
        -\difft{m_{\mathrm{FeO}}} \frac{\Bar{M}_{\mathrm{Fe}}}{\Bar{M}_{\mathrm{FeO}}} \\
    \dot{m}_{\mathrm{O2}, \mathrm{max}} & = \varrho_f \, A_p \, \frac{\mathrm{Sh}\,D_{\mathrm{O2},f}}{d_p}\; \mathrm{ln}(1-Y_{\mathrm{O2},g}) \,, 
    \label{eq:SOLD:diffusion}
\end{align}

where $X_{\mathrm{FeO}}$ specifies the thickness of the oxide layer. Further, the activation temperature, $T_{a,\mathrm{FeO}}$, and pre-exponential factor, $k_{\infty,\mathrm{FeO}}$, are introduced for the oxidation reaction. The evolution of the shrinking iron core is obtained from the mass balance of iron with Eq.~(\ref{eq:SOLD:FeO_curv}). For the total oxygen consumption, the boundary layer diffusion of oxygen represents an upper limit which is described according to Eq.~(\ref{eq:SOLD:diffusion})~\cite{Mi2022}, including Stefan flow.

\subsection{Numerical solution algorithm}
\label{sec:numericalSolution}

Lagrangian particles and Eulerian gas phase are coupled by an operator splitting approach. The gas phase is computed with a hybrid, damped Newton algorithm~\cite{grcar_1988} keeping the heat and mass transfer terms fixed. Thereafter, the particles are advected in a Lagrangian manner from cell to cell through the entire domain while keeping the gas phase conditions fixed. The ODE-solver CVode from the SUNDIALS suite~\cite{hindmarsh_2005}  is used to integrate the particles with respect to the particle time. The mapping between the particle time and the spatial coordinate $x$ follows the trivial transformation $\mathrm{d}x=u_p\mathrm{d}t$. A prerequisite for this approach is that the particle velocity~$u_p$ is larger than zero at all times, which is fulfilled in the one-dimensional simulations of the present work. Aiming for an overall steady state solution, all particles of one size fraction $_k$ can be represented by a single computational particle~\cite{Ravi2022, Filho2018, Hazenberg2021, Messig2017}. The heat and mass transfer for the computational particle is then multiplied by the number flux of the respective particle fraction $k$, $\dot{N}_{p,k}$, which is the number of equal particles passing through the domain per second (unit: \unit{\flux}). The transfer terms to the Eulerian gas phase are then obtained as a sum over all particle fractions with $N_F$ being the total number of particle fractions:
\begin{align}
    \dot{m}_t & = \sum_{k=1}^{N_F} \dot{N}_{p,k} \; \difft{m_{p,k}} \, \frac{1}{u_{p,k}} \label{eq:Mtransfer}, \\
    \dot{h}_t & = \sum_{k=1}^{N_F}  \dot{N}_{p,k} \; \difft{H_{p,k}} \, \frac{1}{u_{p,k}} \label{eq:Qtransfer}, \\
    \dot{Y}_{i,t} & = \dot{m}_t, \; \textbf{if } i=\mathrm{O2}  \label{eq:Ytransfer}.
\end{align}
Using underrelaxation the algorithm described above is iteratively repeated until convergence. Usually, this is achieved within \qty{50} to \qty{250} iterations, depending on the initial conditions.

\subsection{Definition of the equivalence ratio}

The ratio of particles to oxidizer is described by the equivalence ratio~$\phi$. It is defined by the oxygen available in the unburned, premixed inflow, relative to the amount of oxygen~$\dot{m}_{\mathrm{O2},\rm{st}}$ which can be consumed by the particles \cite{Book-Poinsot-NumericalComb}. This stoichiometric mass flux $\dot{m}_{\mathrm{O2},\rm{st}}$~is dependent on the stoichiometric ratio~$s$ in \unit{\kg} oxidizer consumed per \unit{\kg} fuel during full conversion. For all models, the stoichiometric ratio~$s$ is~\num{0.2865}:
\begin{align}
    \phi & = \frac{\dot{m}_{\mathrm{O2},\rm{st}}}{\dot{m}_{\mathrm{O2}}}, \label{eq:EquivalenceRatio} \\
    \dot{m}_{\mathrm{O2},\rm{st}} & = s \; \dot{m}_{\mathrm{Fe}}. \label{eq:stoichiometricRatio}
\end{align}

\subsection{Relevance of discrete effects}

Goroshin and co-workes~\cite{Goroshin2022, Palecka2019, Goroshin1998, Tang2009b, Wright2016} extensively discussed the discrete regime of iron dust flames (and other metal dust flames) which represents a peculiarity not present in gaseous flames. In this discrete regime, the inter-particle heat transfer becomes slower than the combustion of individual particles. In this case, the reaction front speed is governed by the heat conductivity of the gas phase instead of the particle properties. Whether a flame is in the continuous or discrete regime can be determined using the discreteness parameter $\chi = \tau_c \alpha / l_p^2$~\cite{Goroshin1998}. It is the ratio of particle combustion time~$\tau_c$ and the time required to conduct heat over the characteristic distance between two particles~$l_p$. In a suspension with the number density $N_p$, the mean particle distance can be calculated using $l_p = N_p^{-1/3}$. The discreteness parameter $\chi \gg 1$~corresponds to a continuous flame, while in a discrete flame~$\chi \ll 1$. In this study, the discreteness parameter was determined as part of the post-processing for all simulations. It was found to be between \num{2.5} and \num{1100}, especially dependent on the equivalence ratio~$\phi$ and the mean particle diameter. Therefore, all iron dust flames investigated in this work are in the continuous regime.

\section{Verification}
\label{sec:verification}

In this section, the models FIRR, FOSK, and SOLD for single particles are verified by comparison with experimental data for the combustion of individual particles by Ning~et~al.~\cite{Ning2021, Ning2022}. All model parameters and boundary conditions are listed in Tab.~\ref{tab:modelParameters}. If not otherwise stated, the listed values are consistently used for all models and parametric studies in this work. Material properties are obtained from the NIST Chemistry Webbook \cite{NIST_Chase1998}.
\begin{table}
\centering
\caption{Model parameters utilized for the iron particle modeling. Material properties stem from the NIST Webbook~\cite{NIST_Chase1998}.}
\label{tab:modelParameters}
\begin{tabular}{ | r  r  l  l| }
    \hline
    Quantity & Value & Ref. & Model $\vphantom{\bigg|}$ \\
    \hline
    $\rm{Nu, Sh}$ & \num{2} & {\cite{Ranz1952}} & All \\
    $\varrho_{\mathrm{Fe(s)}}$ & $7968-0.334\frac{T_p}{\unit{\kelvin}}$ \unit{\kg\percubicmeter} & \cite{Touloukian1966} & All \\
    $\varrho_{\mathrm{Fe(l)}}$ & $8523-0.8358\frac{T_p}{\unit{\kelvin}}$ \unit{\kg\percubicmeter} & \cite{Kirshenbaum1962} & All \\
    $\varrho_{\mathrm{FeO(s)}}$ & $5776-0.277\frac{T_p}{\unit{\kelvin}}$ \unit{\kg\percubicmeter} & \cite{Touloukian1966, Takeda2009} & All \\
    $\varrho_{\mathrm{FeO(l)}}$ & 4300 \unit{\kg\percubicmeter} & {\cite{Millot2009}} & All \\
    $\epsilon$ & \num{0.7} & {\cite{VDI-Heat-Atlas-2010}} & All \\
    $s$ & \num{0.2865} & {} & All \\
    $T_{\rm{ign}}$ & \qty{1000}{\kelvin} & {} & FIRR \\
    $\tau_{c}$ & $0.038 \frac{d_{p,0}}{\unit{\mum}}^{1.63}\unit{\milli\second}$ & {\cite{Ning2021}} & FIRR \\
    $k_{\infty}$ & \qty{7500000}{\meter\persecond} & {\cite{Hazenberg2021}} & FOSK \\
    $k_{\infty,\mathrm{FeO}}$ & \qty{2.670e-4}{\squaremeter\persecond} & {\cite{Mi2022}} & SOLD \\
    $T_{a}$ & \qty{14400}{\kelvin} & {\cite{Hazenberg2021}} & FOSK \\
    $T_{a,\mathrm{FeO}}$ & \qty{20319}{\kelvin} & {\cite{Mi2022}} & SOLD \\
    \hline
\end{tabular}
\end{table}

\begin{figure}[h]
    \centering
    \includegraphics[scale=1]{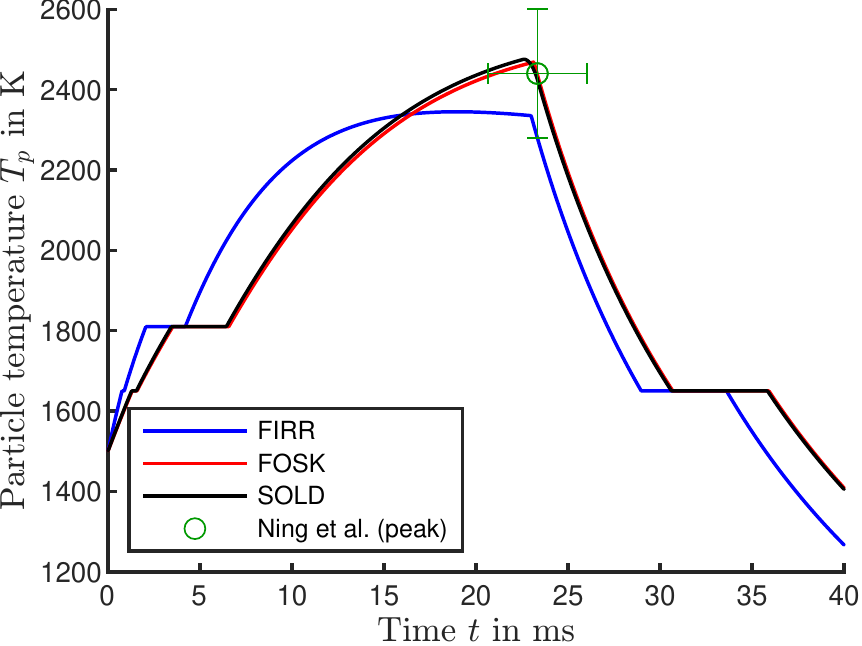}
    \captionsetup{justification=centering}
    \caption{Particle temperature profiles for laser-heated single particles with \qty{54}{\mum} diameter in air at \qty{300}{\kelvin}. For orientation, the green marker shows the averaged peak temperature measurements of Ning~et~al.~\cite{Ning2021, Ning2022}.}
    \label{fig:singleParticle}
\end{figure}

In their experiments, Ning~et~al.~\cite{Ning2021, Ning2022} used particles with diameters from \qty{26}{\mum} to \qty{54}{\mum}, which were laser-heated above their ignition temperature in different oxygen-nitrogen mixtures. For an exemplary \qty{54}{\mum} particle in air, Fig.~\ref{fig:singleParticle} shows the calculated particle temperatures over time for the three models, respectively. An initial temperature of \qty{1500}{K} was chosen in our simulation, since this temperature assures ignition for all atmospheres and particle sizes, but is still beneath the melting temperature. This value has also been chosen by other authors~\cite{Thijs2022a}. For comparison with the experimental results, the green marker shows the mean combustion time of~\qty{21}{\milli\second} and mean peak temperature of~\qty{2440}{\kelvin} measured by Ning~et~al.~\cite{Ning2021, Ning2022}. The error bars indicate the standard deviations reported in their work.
As observed in Fig.~\ref{fig:singleParticle}, the calculated peak temperatures and times match the experimental results by Ning~et~al.~\cite{Ning2021, Ning2022} within the range of experimental uncertainty, i.~e.~deviations of \qty{1}{\%} for time-to-peak and \qty{3.8}{\%} for temperature.
Fig.~\ref{fig:singleParticle} illustrates that the three particle models, i.~e.~FIRR and FOSK, yield almost equal temperature evolution results with respect to peak temperatures and burnout times for single particle burnout. For the models FOSK and SOLD, model parameters have been obtained from the recent literature as indicated in Tab.~\ref{tab:modelParameters}. Specifically for the FIRR model, the combustion time is a free parameter and is calibrated to \qty{23}{\milli\second} for \qty{54}{\mum}~particles in order to match the burnout times obtained with the models FOSK and SOLD. As observed in Fig.~\ref{fig:singleParticle}, this further entails a peak temperature comparable to the models FOSK and SOLD. 

\begin{figure}[h]
    \centering
    \includegraphics[scale=1]{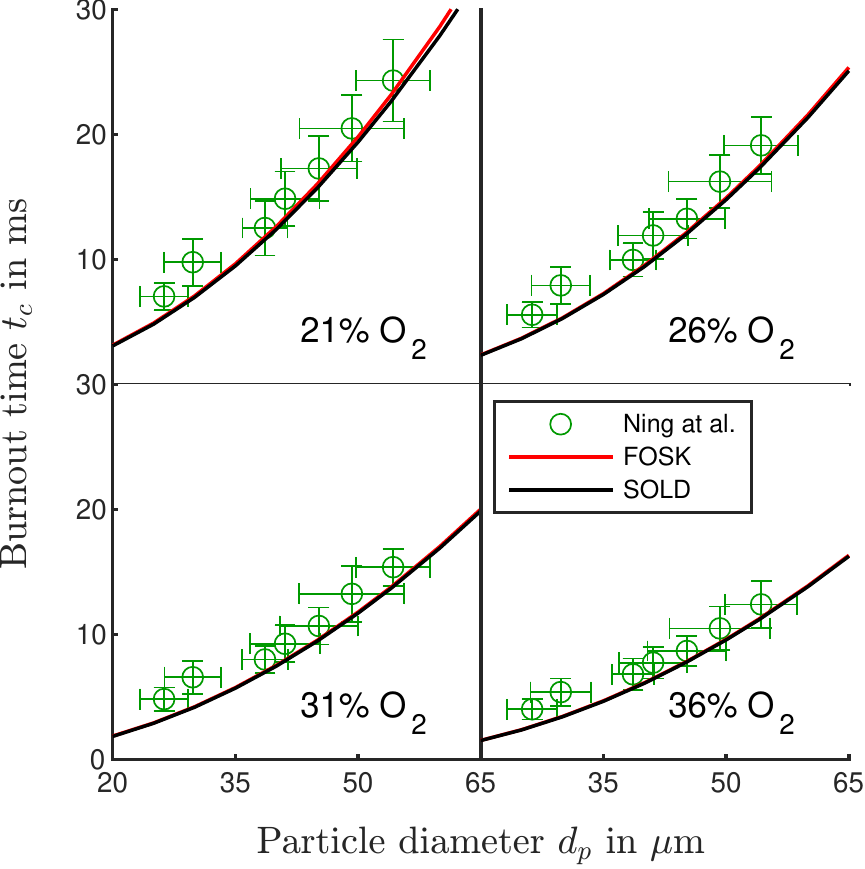}
    \captionsetup{justification=centering}
    \caption{Burnout times for laser-heated single particles with different diameters in different oxygen-nitrogen mixtures at \qty{300}{\kelvin}. The green markers show the experimental results of Ning~et~al.~\cite{Ning2021}. The FIRR model is not included here, since the burnout time is a free parameter for this model.}
    \label{fig:singleParticleTimes}
\end{figure}

A further validation using the experimental burnout times for the other diameters and atmospheres investigated by Ning~et~al.~\cite{Ning2021} is shown in Fig.~\ref{fig:singleParticleTimes}. The calculated values show a good agreement with the experimental data in the entire range of diameters and atmospheres.

The first comparison between the three models with a consistent set of model parameters reveals that FOSK and SOLD show a similar diffusive burnout. Note that, despite the increasing model complexity from FIRR to FOSK to SOLD, FOSK and SOLD use the same approach for the diffusion limitation, which governs most of the high temperature oxidation of the particles. 
Overall, the peak temperatures and burnout times of all three models agree reasonably well with the experimental results and are therefore suitable to elucidate the dependency between single particle and reaction front characteristics. In the next section, we test whether the calibration of single particle burnout behavior is sufficient to yield similar predictions for reaction front speeds and structures.

\section{Results and Discussion}
\label{sec:resultsAndDiscussion}

Using the three particle models in the planar flame configuration shown in Fig.~\ref{fig:setup}, the reaction front speed~$s_f$ can be calculated for iron-air suspensions prescribing arbitrary particle size distributions~(PSDs). First, a macroscopic quantification of the influence of the particle modeling and particle size distribution on the reaction front speed is presented. To this end, we systematically increase the complexity of the PSDs from bidisperse (Sec.~\ref{sec:bidisperse}) to generic, polydisperse  iron-air suspensions (Sec.~\ref{sec:genericPolydisperse}). Thereafter, the microscopic structure of the reaction fronts in the polydisperse iron-air suspensions is analysed in Sec.~\ref{sec:sequentialRunaway}. Eventually, the reaction front speeds obtained for highly resolved and gradually coarsened PSDs of real iron powder samples are considered and discussed in Sec.~\ref{sec:realisticPolydisperse}.

\subsection{Bidisperse iron-air suspensions}
\label{sec:bidisperse}

\begin{figure}[htbp]
    \centering
    \includegraphics[scale=1]{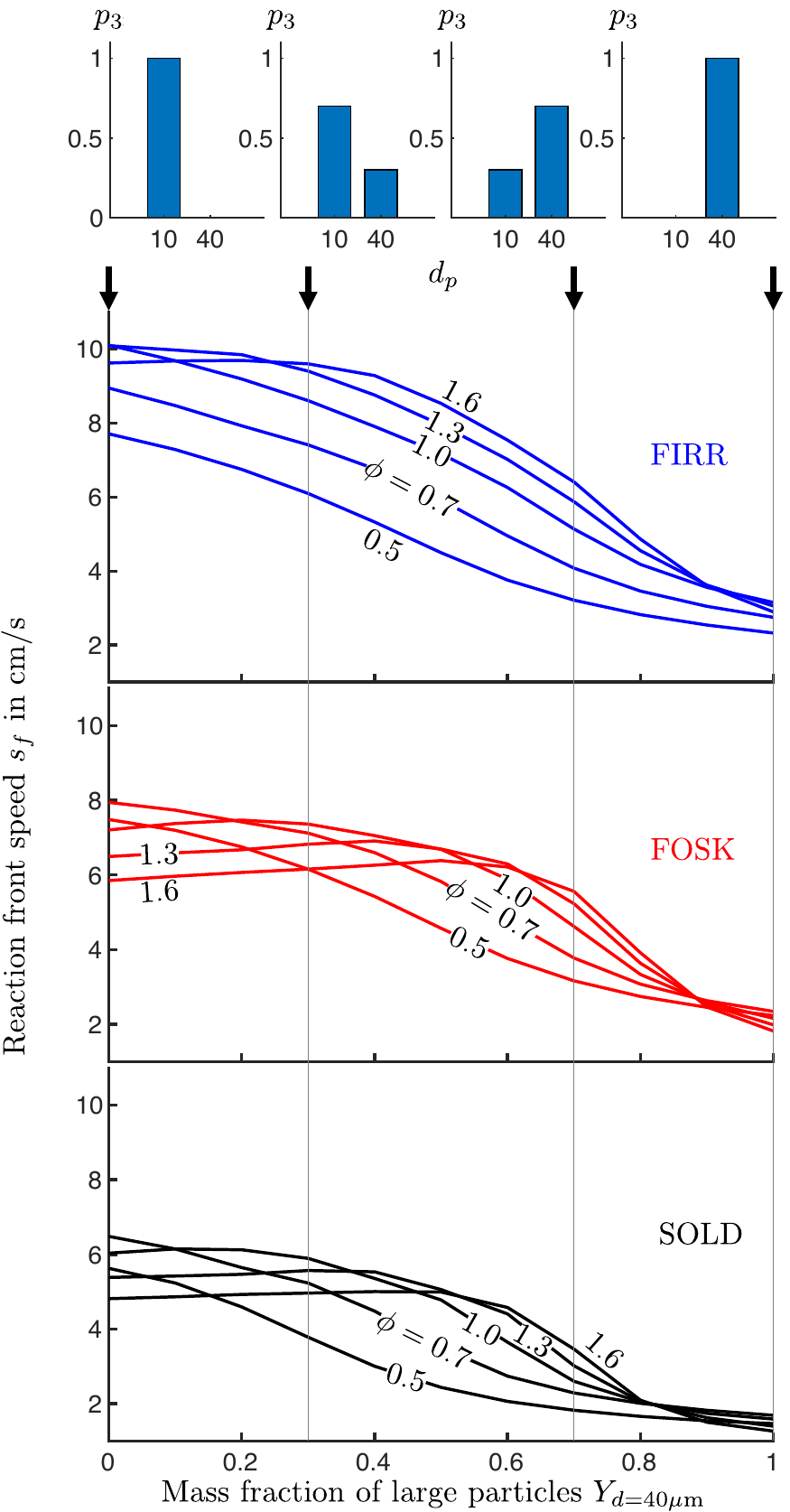}
    \captionsetup{justification=centering}
    \caption{Calculated reaction front speeds for bidisperse iron powder suspensions in air. From the left to the right, the PSD is varied as indicated by the bar diagrams on the top. The different particle models are shown above each other in different colors. Lines of equal color represent a variation of the equivalence ratio.}
    \label{fig:bidisperse}
\end{figure}

The investigation performed in this section is based on the generic study by Goroshin~et~al.~\cite{Goroshin2000} on bidisperse iron-air suspensions, which was later numerically investigated by Ravi~et~al.~\cite{Ravi2022}. In our approach, we consider iron powder which consists of two particle fractions with diameters of~\qty{10}{\mum} and \qty{40}{\mum}. Fig.~\ref{fig:bidisperse} shows the calculated reaction front speeds for the bidisperse iron-air suspensions. The ratio of small particles to large particles is varied, starting with monodisperse small particles which refers to the left boundary position in Fig.~\ref{fig:bidisperse}. Successively, small particles are replaced by the larger particles (with equivalent mass) until eventually the powder only consists of large particles. Hence, the right boundary position in Fig.~\ref{fig:bidisperse} again refers to a monodisperse iron-air suspension. For orientation, the transition of the PSD is indicated by the plots on the top of Fig.~\ref{fig:bidisperse}, showing the mass fraction $p_3$ of each particle size fraction at exemplary positions. The parameter study is performed for multiple equivalence ratios~$\phi$ utilizing all three models, FIRR, FOSK, and SOLD. 

First, the obtained reaction front speeds for the two monodisperse cases on the left and right boundaries in Fig.~\ref{fig:bidisperse} are compared. Interestingly, the three models yield significantly different reaction front speeds despite the good agreement for the single particle conversion (cf. Fig.~\ref{fig:singleParticle}). For monodisperse setups with only \qty{10}{\mum} particles, SOLD leads to the lowest reaction front speeds of between \qty{4.8}{\centi\meter\persecond} and \qty{6.5}{\centi\meter\persecond}. With the FOSK model, the reaction fronts are predicted to be \qty{23}{\%} faster on average with \qty{5.9}{\centi\meter\persecond} to \qty{7.9}{\centi\meter\persecond}. Utilizing the FIRR model results in even higher reaction front speeds. For the monodisperse flames with \qty{40}{\mum} particles, the absolute flame speeds predicted by all models are smaller but with identical relative relations. This clearly shows that a model calibration with combustion times and peak temperatures obtained from single particle experiments is insufficient for a consistent prediction of the reaction front speed. Hence, differences in the particle kinetics, which show only small effects on the single particle combustion, have a major influence when considering reaction fronts in metal particle-air suspensions. 

For the bidisperse particle size distributions in the middle section of Fig.~\ref{fig:bidisperse}, all three models qualitatively show similar trends: The computed reaction front speeds decrease in a non-linear manner with an increase of the mass of large particles in the powder composition, and the trend depends on the equivalence ratio. An increase of the equivalence ratio also leads to an increase of the reaction front speed in most cases. However, for powders with a sufficient mass fraction of small particles this trend reverses and the point of reversal is model specific.
This considerable non-linear response of the reaction front speed to the PSD and the equivalence ratio prohibits its approximation with a simple average particle size, which is the answer to \QavgDiaPoss{}. 

Lean cases ($\phi < 1$) show a steep decrease in the reaction front speed on the left side, but a flattening on the right side. This shows that adding a few large particles to a lean flame, mainly driven by small particles, directly leads to a drop in the reaction front speed. However, adding a few small particles in a flame of mainly large particles has only a minor effect. Hence, large particle fractions show an increased influence on the total reaction front speed in lean flames. The described shape was also found by Goroshin~et~al.~\cite{Goroshin2000} based on their analytical model. One of their model assumptions was, that the flame is very lean such that the oxygen partial pressure in the flame is not affected by the combustion. Similarly to the present work, they concluded that there is no average particle size which can be used to approximate the flame speed in a lean, binary fuel mixture. \cite{Goroshin2000}

Notably, the described trend completely reverses in rich flames ($\phi > 1$, cf. Fig.~\ref{fig:bidisperse}). Here a plateau is found on the left while a steep decrease in~$s_f$ can be observed more on the right of the graph. A similar curve was reported by Ravi~et~al.~\cite{Ravi2022} for a stoichiometric flame with a Model similar to FOSK. This means that including a few large particles to lean flame of mainly small particles does not lead to a drop in the reaction front speed. The drop only occurs when enough large particles are included to reach a threshold after which $s_f$~suddenly decreases. The previous studies explained this behavior by the separation of flame fronts~\cite{Goroshin2000, Ravi2022}, which is examined in further detail in Sec.~\ref{sec:sequentialRunaway}. On the other hand, including a few small particles in a flame of mainly large particles directly promotes the reaction front speed. Thus, the presence of small particle fractions show to have a strong influence on the total reaction front speed in rich flames.

Considering the results discussed in this section, at this point an answer to \QimpBouCond{} can already be provided: The PSD is a crucial parameter which needs to be specified when investigating reaction front speeds of iron-air suspensions (numerically or experimentally). With this background, a secondary related question question arises: what is the required resolution of the PSD to reliably predict the reaction front speed? Another observation worth mentioning is that the maximum reaction front speed is found at different equivalence ratios depending on the particle size distribution as well as the particle model.

\begin{figure}[h]
    \centering
    \includegraphics[scale=1]{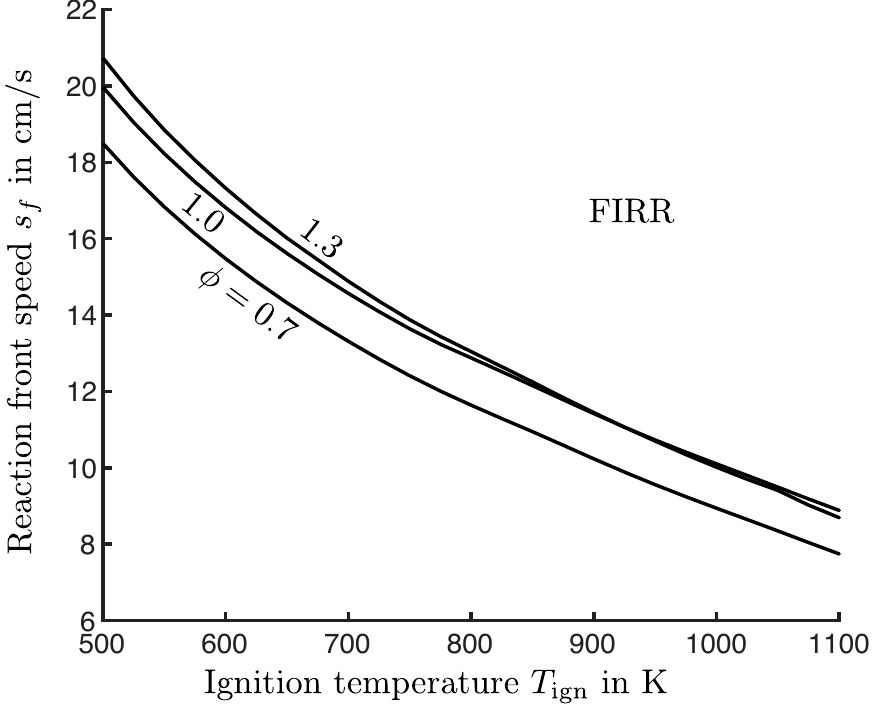}
    \captionsetup{justification=centering}
    \caption{Dependency of the reaction front speed on the ignition behavior of the particle model.}
    \label{fig:sLoverTign}
\end{figure}

Again, it is interesting to note that the strong model dependency of the calculated reaction front speeds is not reflected in the single particle calculations in Sec.~\ref{sec:verification}. An influential model characteristic, which can at least partially explain the differences in~$s_f$, is the ignition temperature~$T_i$. Based on their analytical model for particle flames, Goroshin~et~al.~\cite{Goroshin1998} already pointed out this strong dependency:
\begin{align}
    s_f \propto \sqrt{\frac{T_\infty-T_i}{T_i-T_0}},
\end{align}
where $T_{\infty}$ is the adiabatic flame temperature. For illustration, the ignition temperature, which is a free model parameter for the FIRR model, is varied and the calculated reaction front speeds for a monodisperse setup are shown in Fig.~\ref{fig:sLoverTign}. The decrease of the reaction front speed with increasing ignition temperature is recovered by our numerical model and reconfirms the theoretical prediction of Goroshin~et~al.~\cite{Goroshin1998}. Thereby, at least a partial answer is given to questions \QmostImpAsp{} and \QfurExpReq{}: The combustion time and peak temperatures are key observables but additional information is required for the determination of reaction front speeds. Specifically, the ignition temperature of the particles needs to be taken into account, for which currently only insufficient experimental data is available.

\subsection{Generic Polydisperse Flame}
\label{sec:genericPolydisperse}

\begin{figure}[htbp]
    \centering
    \includegraphics[scale=1]{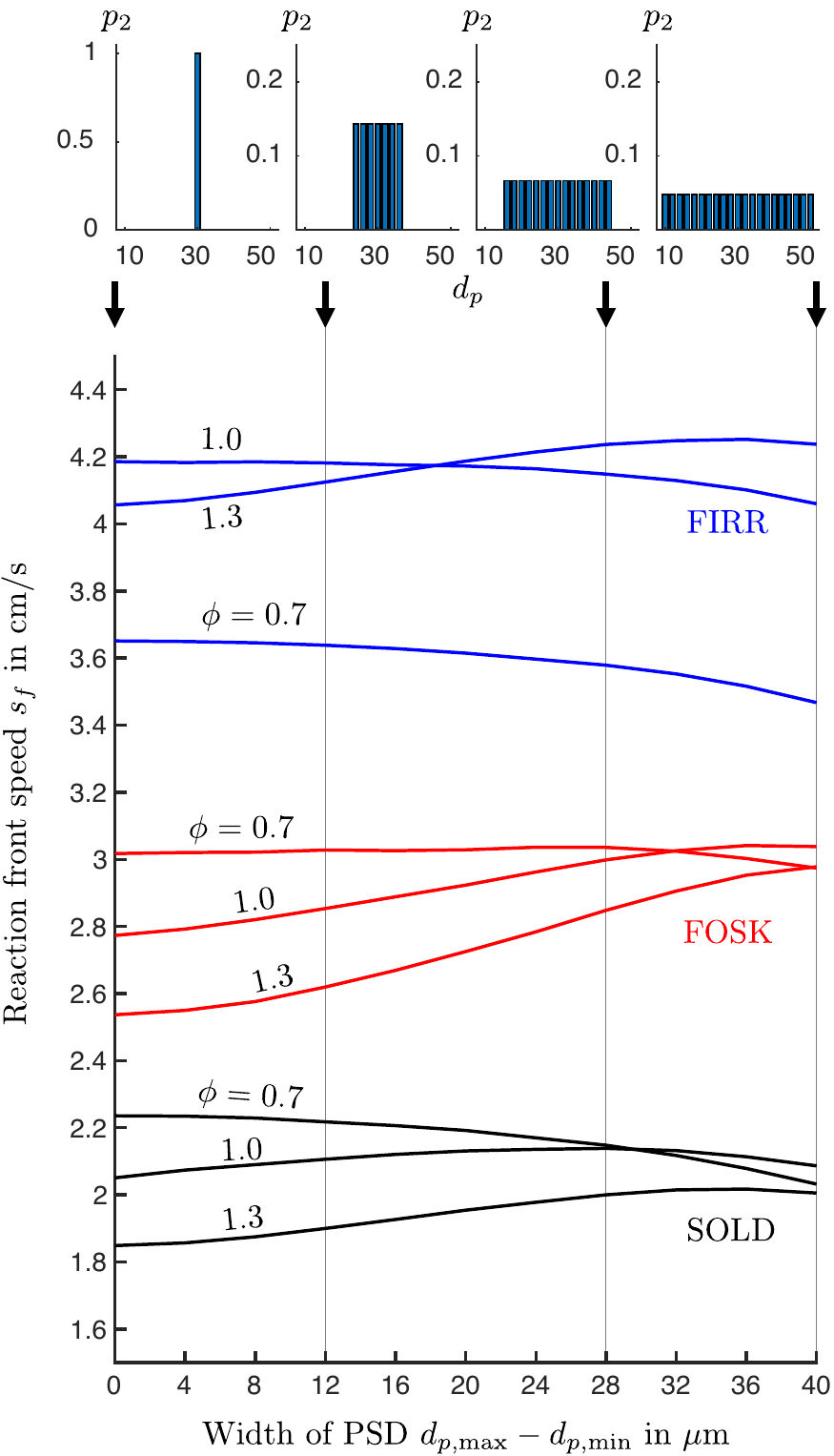}
    \captionsetup{justification=centering}
    \caption{Reaction front speed $s_f$, dependent on the PSD in a generic polydisperse iron-air suspension. The number of particle fractions increases from one (monodisperse) on the left to 21 on the right. The Sauter mean diameter~$d_{32}$ is fixed to \qty{30}{\mum} in all cases.}
    \label{fig:rectangle}
\end{figure}

Proceeding to polydisperse iron-air suspensions, we first prescribe a generic PSD allowing for a systematic analysis of polydispersity effects on the reaction front speed, which is shown in Fig.~\ref{fig:rectangle}. The PSD is chosen such that each particle fraction contributes with the same reactive surface area to the total reactive surface area of the iron powder and the Sauter mean diameter~$d_{32}$ remains fixed at \qty{30}{\mum} for all cases. Hence, the PSD exhibits a box-like shape when considering the $p_2$ probability distribution. This artificial distribution is designed to evaluate whether the respective contributions to the overall reaction font speed by the small and large particle fractions level out, such that the PSD can be simplified by a monodisperse equivalent (\QavgDiaPoss).

The starting point of the parameter study is a delta peak, i.~e.~monodisperse particles of~\qty{30}{\mum} diameter, which is shown on the top left of Fig.~\ref{fig:rectangle}. Subsequently, from top left to top right in Fig.~\ref{fig:rectangle}, particle fractions are added on both sides to this reference diameter~\qty{30}{\mum}. The diameter difference of two neighbouring particle fractions is~\qty{2}{\mum}. In the considered cases, it is found that the reaction zone of each particle fraction overlaps with the one of the neighbouring particle fractions inside of the reaction front and no reaction front separation occurs. The resulting PSDs are distinguished by their width, which refers to the diameter difference between the largest particle fraction and the smallest particle fraction. In the PSD with the highest width investigated in this study, the particle diameters range from \qty{10}{\mum} to \qty{50}{\mum} (shown in the top right in Fig.~\ref{fig:rectangle}). Thus, while keeping a constant Sauter mean diameter, the polydispersity increases from left to right in Fig.~\ref{fig:rectangle} which is achieved by adding proportionally smaller and larger particle fractions (referring to their surface area).

Consistently with the results from the bidisperse setup, the FIRR model yields the highest reaction front speeds and the SOLD model the lowest. Independently from the model, and with increasing width of the PSD (moving from left to right in Fig.~\ref{fig:rectangle}), the reaction front speed decreases for lean cases and initially increases for rich cases. 
Since the small particles predominantly influence the reaction front speeds in fuel-rich iron-air suspensions, they overcompensate the influence of the additional large particles and the reaction front speed increases. In the lean case, the larger particles' thermal inertia overcompensates the promoting effect of small particles and the reaction front speed decreases. The equivalence ratio, at which the transition between promoting or impeding effect of polydispersity occurs, is again model specific. This is observed from the stoichiometric cases ($\phi=1$) for the models FIRR, FOSK, and SOLD which show different trends. Hence, the contribution of a particle size fraction to the overall flame behavior is not only dependent on the PSD, but also on the particle model, more precisely the model of the particle kinetics. 
The significant deviations for the reaction front speed obtained from utilizing different particle models demands more detailed investigations of the kinetically limited regime and warrants novel experiments addressing this aspect.
Further, the diverse behavior at constant Sauter mean diameter again underlines that there is no representative monodisperse particle fraction with some sort of average diameter which could be used to approximate the reaction front characteristics of polydisperse iron-air suspensions.

\subsection{Reaction front structure and sequential thermal runaway}
\label{sec:sequentialRunaway}

\begin{figure}[h]
    \centering
    \includegraphics[scale=1]{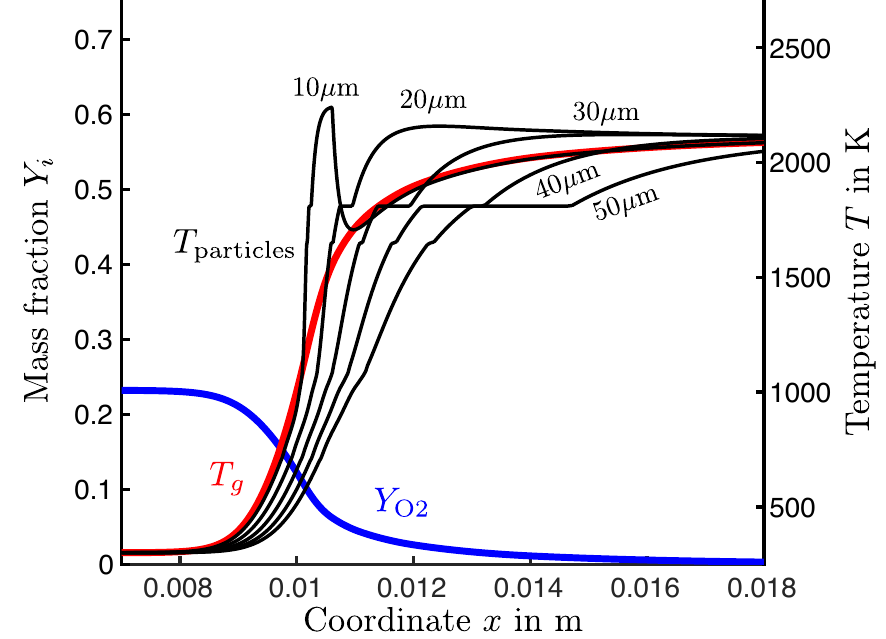}
    \captionsetup{justification=centering}
    \caption{Flame structure of a generic polydisperse flame with $\phi=1.3$ (FOSK model). The particles' thermal runaways occur sequentially in the reaction front (thin black lines) at different local combustion environments (thick lines).}
    \label{fig:depictSequence}
\end{figure}

To understand the previously discussed behavior, we analyze an exemplary reaction front for a polydisperse iron-air suspension in detail. Figure~\ref{fig:depictSequence} shows the reaction front structure for a case from section~\ref{sec:genericPolydisperse} with $\phi=1.3$ employing the FOSK model. The PSD consists of~\num{21} particle fractions with diameters from \qty{10}{\mum} to \qty{50}{\mum} in steps of~\qty{2}{\mum}. The particle temperature profiles are shown as a function of the spatial coordinate~$x$ for every fifth particle fraction as thin black lines in Fig.~\ref{fig:depictSequence}. The thicker solid lines show the gas phase temperature (red) and oxygen mass fraction (blue) across the reaction front. The thermal runaways of the particles are clearly visible as peaks in the particle temperature. Small particles undergo their thermal runaway more upstream, while the larger particles undergo it more downstream. The individual particle reaction zones overlap and form a single, collective reaction front with a width of around \qty{10}{\milli\meter}. In other simulations, we have found reaction front widths of up to \qty{30}{\milli\meter}, especially for stoichiometric iron-air suspensions. The different slopes of the particle temperatures in the preheating zone show that this effect is mainly due to the thermal inertia of the particles. Since the smaller particles have a higher specific surface area, they heat up faster than the large ones, resulting in a 'sequence' of thermal runaways.

\begin{figure}[h]
    \centering
    \includegraphics[scale=1]{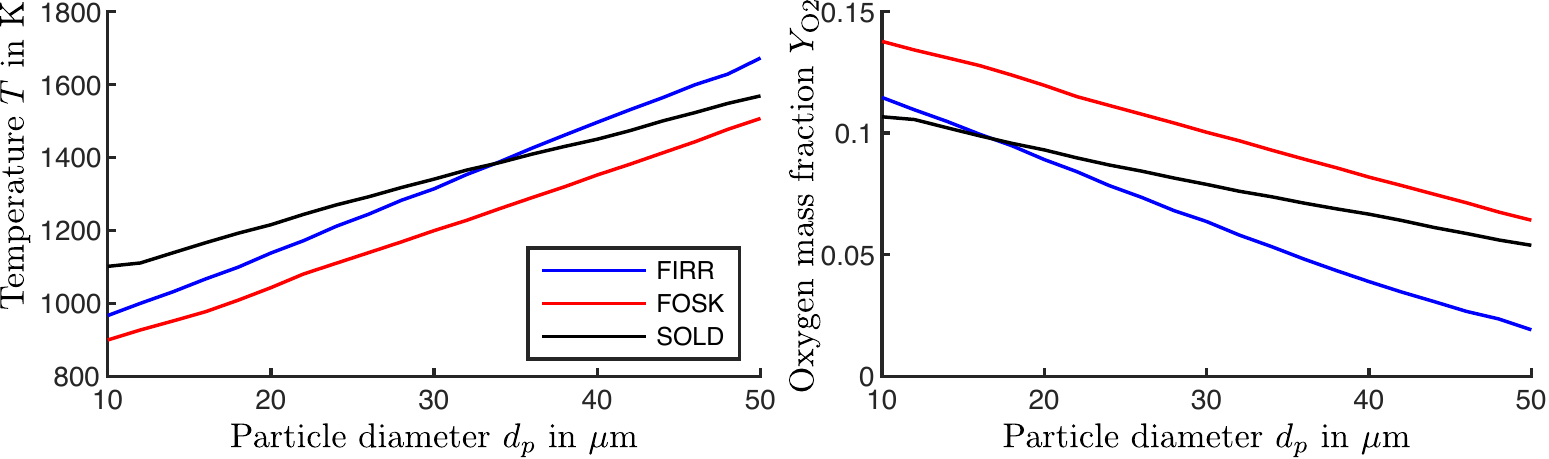}
    \captionsetup{justification=centering}
    \caption{Gas phase environment for differently sized particle fractions during their thermal runaway in a generic, polydisperse, rich flame ($\phi=1.3$, the PSD consists of~\num{21} particle fractions with diameters from \qty{10}{\mum} to \qty{50}{\mum}). Each particle fraction burns in an individual, unique environment which is also influenced by the particle model.}
    \label{fig:ignitionConditions}
\end{figure}

As depicted in Fig.~\ref{fig:depictSequence}, the oxygen content in the gas ($Y_{\mathrm{O2}}$, blue line) decreases in the reaction front while the temperature ($T_g$, red line) increases. For every particle fraction, the position of the thermal runaway in the reaction front thereby determines its local combustion environment (ambient temperature, oxygen availability). Hence, the sequential thermal runaway which stems from the different thermal inertia causes each particle fraction to burn in its own, unique environment. This is illustrated for each particle fraction in Fig.~\ref{fig:ignitionConditions}. The figure shows the temperature (left) and oxygen mass fraction (right) of the gas phase over the particle diameter at the characteristic point, where the particle temperature exceeds the gas phase temperature, i.~e.~an indicator for the thermal runaway. The three colors distinguish the different particle models.

As to be expected from the sequential thermal runaway, smaller particles burn upstream in a more oxygen-rich environment than the larger particle fractions. Note that the local oxygen mass fraction during the thermal runaway in the flame (\qty{2}{\%} to \qty{14}{\%}) is well below the oxygen mass fractions studied in many recent single particle combustion experiments (\qty{15}{\%} to \qty{40}{\%})~\cite{Ning2021, Ning2022, Li2022}. In general, this difference leads to increased particle combustion times in iron-dust flames compared to single particle experiments. A counteracting but secondary effect promotes the particle oxidation due to the elevated gas phase temperatures downstream, where larger particles exhibit a thermal runaway. Thus, the diffusion of oxygen through the particle boundary layer is accelerated, which determines the conversion rate in the diffusion-limited regime. For the rich flame which is depicted in Fig.~\ref{fig:depictSequence}, the small particle fractions show an extensive thermal overshoot exceeding the gas phase temperature. The relationship between particle peak temperature and flame temperature was previously theoretically analysed by Tang~et~al.~\cite{Tang2011}. For monodisperse lean iron-air suspensions, the authors predicted an intense thermal overshoot similar to the small particle fractions shown in Fig.~\ref{fig:depictSequence}. On the other hand, they stated that the particle peak temperature is limited to the flame temperature in monodisperse rich flames due to oxygen depletion. With respect to the rich cases, polydisperse iron-air suspensions require a more differentiated perspective: due to the different individual combustion environments which depend on the PSD, small particles still burn under excess oxygen even in rich flames and exhibit a thermal overshoot, since they do not compete with the large particles for the oxygen. The amount of excess oxygen decreases with increasing particle diameter until the largest particles finally burn under the expected oxygen depleted conditions. In relation to question \QaddEffPoly{}, it is an additional effect of polydispersity that smaller particles burn upstream in leaner conditions than the 'flame average' while large particles burn downstream in a richer environment than the 'flame average'.

Figure~\ref{fig:ignitionConditions} further shows differences in the individual particle combustion environment originating from the different particle models. In this particular case, an exchange of the particle model leads to a variation of the local combustion environment by up to \qty{200}{\kelvin} and \qty{4.5}{\%} oxygen mass fraction, dependent on the particle size. Clearly, this is an effect of the particle kinetics, since the models FOSK and SOLD use the same diffusion model. The main difference is the ignition behavior of the models. For the FIRR model, the thermal runaway is initiated when the particle reaches \qty{1000}{\kelvin} without any further condition depending on the particle's environment or its size. For the FOSK model, particle size and surrounding oxygen influence the occurrence of the thermal runaway and it has been shown that the ignition temperature is inverse to the particle diameter~\cite{Mi2022, Hazenberg2021}. For the SOLD model, the ignition temperature is a complex function of the particle size and the formation of the oxide layer, involving saturation effects~\cite{Mi2022}. These different ignition behaviors significantly contribute to the ignition delay time and the position of the thermal runaway in a reaction front. As a result, even if the particle burnout is governed only by the diffusion of oxygen from the bulk to the particle (diffusion-limited regime), the particle kinetics play an important role for reaction front propagation. In fact, the particle kinetics and ignition behavior set the boundary conditions for the burnout limited by external diffusion and must be well modeled. This is an important answer to \QmostImpAsp{}. The different slopes for the different models indicate that the ignition temperature is not a fixed value, but a complex function of the particle kinetics and the surrounding gas phase, as already discussed by Goroshin~et~al.~\cite{Goroshin2022}.

The previously discussed mechanisms in polydisperse reaction fronts are strongly coupled. The position of the thermal runaway for an individual particle fraction within the flame front is a function of the entire PSD and the particle kinetics. It further depends on the oxygen and temperature profile in the reaction front~\cite{Goroshin2022}. The latter set the boundary conditions for the particle's diffusion-limited burnout and its heat release, which again influence the gas phase conditions in a non-linear two-way coupling. Especially oxygen, which is consumed by small particles upstream, is not available for the large particles downstream, directly affecting their combustion times and heat release rate. These complex relationships show that the prediction of the behavior of real iron-dust flames requires further information beyond single particle experiments and experiments with nearly monodisperse metal powders (even though these experiments deliver important reference data for model development and calibration). For a comprehensive and predictive modeling approach, the PSD needs to be considered in a coupled approach with the individual particle combustion times being part of the solution~\cite{Goroshin2022}.

\subsection{Realistic Polydisperse Flame}
\label{sec:realisticPolydisperse}

\begin{figure}[htb]
    \centering
    \includegraphics[width=6cm]{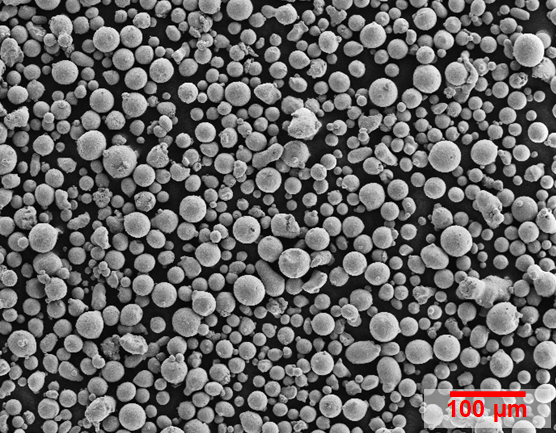}
    \captionsetup{justification=centering}
    \caption{Image of the iron powder using a scanning electron microscope - SEM (secondary electron contrast, Zeiss Gemini Leo 1530)}
    \label{fig:SEM}
\end{figure}

\begin{figure}[htb]
    \centering
    \includegraphics[scale=1]{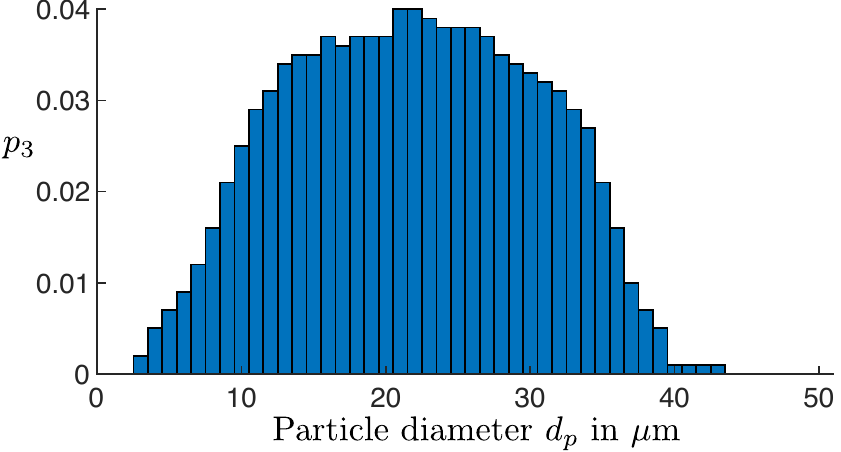}
    \captionsetup{justification=centering}
    \caption{Probability density function of the measured PSD, recorded from real iron powders using a Camsizer X2 (Microtrac Retsch GmbH, Germany, sample size approx. 5 million particles) with a resolution of \qty{1}{\mum}.}
    \label{fig:realPSDs_PSDs}
\end{figure}

From the results discussed in the previous sections it is obvious that the PSD is an essential parameter in iron dust flames which needs to be considered for predictive numerical simulations. One remaining question is: Which resolution viz. particle size classification is necessary for the PSD to describe the reaction front speed with sufficient accuracy? To find an answer to this question, exemplary calculations are carried out prescribing a realistic PSD measured for a real iron powder (sieved at around \qty{45}{\mum} mesh size) produced by Eckart TLS GmbH in Germany. The iron powder has been produced by inert gas atomization (Argon) and a microscopical image of the material is shown in Fig.~\ref{fig:SEM}, revealing that the particles are predominantly spherical in shape. A Camsizer X2 device (Microtrac Retsch GmbH, Germany) has been used to measure the particle widths and, thus, to obtain the PSD (approximately 5 million particles were considered) with a class width of~\qty{1}{\mum} in the range from \qty{1}{\mum}~to~\qty{100}{\mum} diameter. For the principle of the Camsizer X2 (dynamic image analysis), the reader is referred to the literature~\cite{Westermann2016}. The smallest and largest particles detected have a width of~\qty{2}{\mum} and~\qty{43}{\mum}, respectively ($d_{10} = \qty{10.1}{\mum}$, $d_{50} = \qty{21.4}{\mum}$, $d_{90} = \qty{32.6}{\mum}$). The fully resolved PSD with \num{41} particle size fractions is shown in Fig.~\ref{fig:realPSDs_PSDs}. With this PSD, the reaction front speeds $s_f$ are calculated with the models FOSK and SOLD for different equivalence ratios $\phi$.

\begin{figure}[htb]
    \centering
    \includegraphics[scale=1]{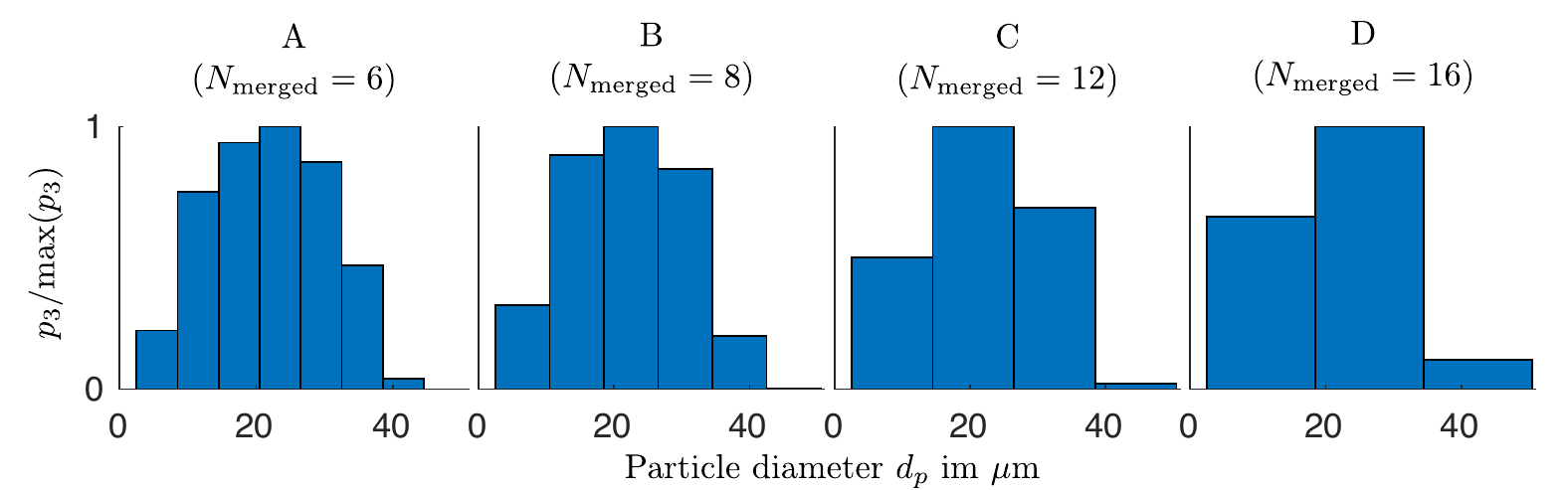}
    \captionsetup{justification=centering}
    \caption{PSDs with decreasing resolution, derived from a measured PSD by merging variable numbers of particle size fractions. The density probability is normalized for better comparison.}
    \label{fig:mergedPSDs}
\end{figure}

In order to investigate the influence of the resolution, the PSD is successively coarsened by merging particle size fractions (new class widths \qty{6}{\mum} to \qty{16}{\mum}, cf. Fig.~\ref{fig:mergedPSDs}). Therefore, a variable number of adjacent particle size fractions,  $N_{\rm{merged}}$, are combined to one merged particle size fraction. With an increasing number of original size fractions forming one merged size fraction, the resolution of the PSD decreases.

The corresponding diameters of the merged size fractions are calculated as the mean diameter of all original size fractions which it replaces. The mass weighted probability density $p_3$ is calculated by the summation of the mass weighted probability densities of the respective original size fractions. This means that the total mass of the original size fractions is preserved during the merge.

\begin{align}
    d_{\rm{merged}} & =    \sum_{s=1}^{N_{\rm{merged}}}{d_s} \frac{1}{N_{\rm{merged}}}, \\
    p_{3,\rm{merged}} & =  \sum_{s=1}^{N_{\rm{merged}}}{p_{3,s}}.
\end{align}

\begin{figure}[htb]
    \centering
    \includegraphics[scale=1]{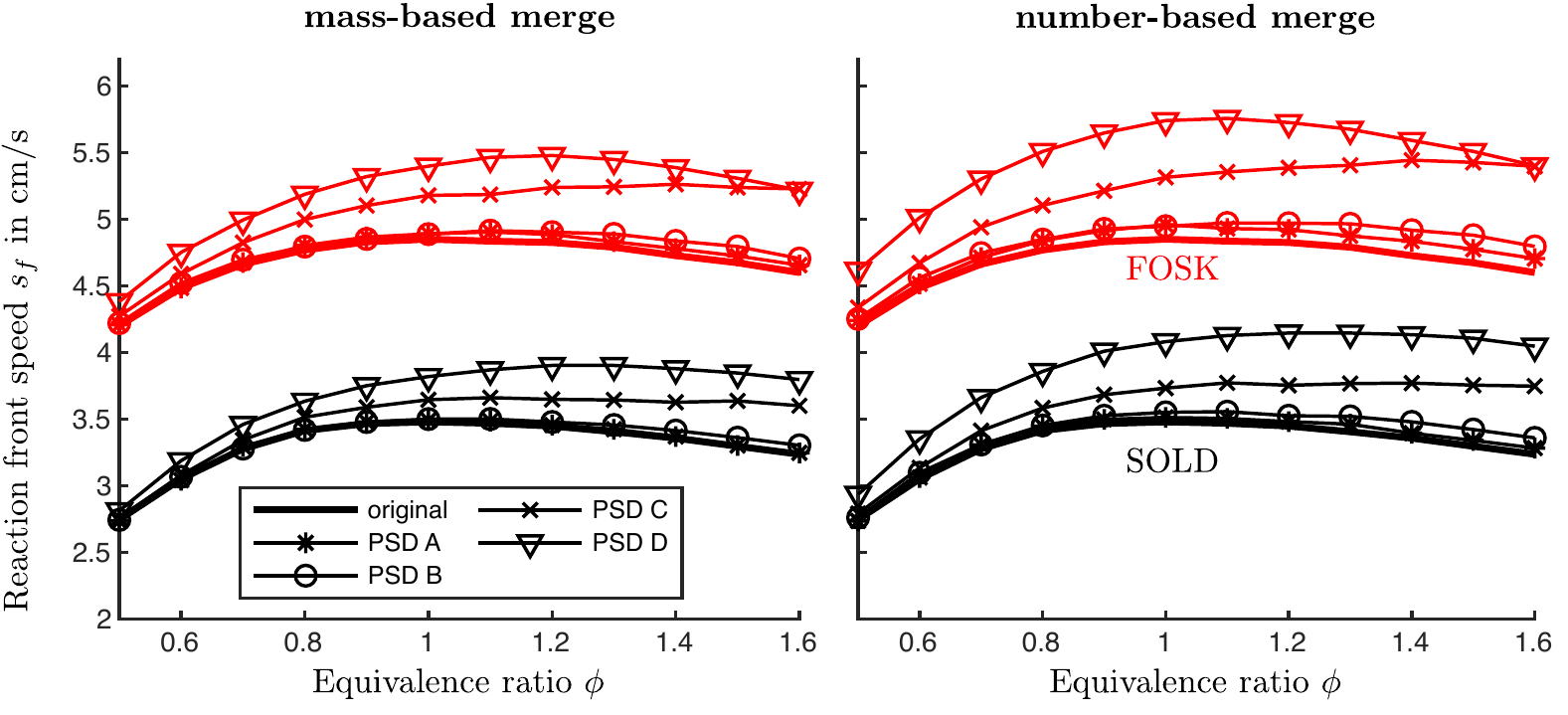}
    \captionsetup{justification=centering}
    \caption{Reaction front speeds for iron-air suspensions calculated using the measured PSDs of real iron powder with successive coarsening of the PSD (see also: Figs.~\ref{fig:realPSDs_PSDs} and \ref{fig:mergedPSDs}).}
    \label{fig:realPSDs}
\end{figure}

The number of particle size fractions which are respectively merged to one new size increases from A to D, see Fig.~\ref{fig:mergedPSDs}. The remaining number of particle size fractions in these PSDs are 7, 6, 4 and 3, respectively. The simulation is repeated with these new PSDs and the results are shown on the left side of Fig.~\ref{fig:realPSDs}.

With decreasing resolution, the calculated reaction front speeds increase compared to the calculations with the original PSD. The difference is more significant in rich flames than in lean flames. This can be explained by the increased contribution of small particles to the reaction front speed in rich flames, as described in the previous sections. It is safe to assume that the reduction in resolution has a particular effect on these small particle fractions. A precise measurement with a comparable class width for different application scenarios of real powder feedstocks is of high importance for understanding the interrelations between the applied material and corresponding combustion behavior.

The relative difference of the reaction front speed to the original calculation remains below~\qty{1.5}{\%} when merging up to 6 particle size fractions (bin size: \qty{6}{\mum}). Despite the fact that this value is expected to vary with the shape of the original PSD, this shows that the merging of particle fractions into a coarser discretisation of the PSD is a practicable technique to save computational costs while keeping a reasonable accuracy for the reaction front speed predictions.

PSDs are reported in the literature using different units. For comparison, also an alternative merging strategy is used for which the number of particles is preserved. Thus, the \emph{number}-weighted probability densities $p_0$ are summed up when merging particle fractions
\begin{align}
    p_{0,\rm{merged}} & =  \sum_{s=1}^{N_{\rm{merged}}}{p_{0,s}},
\end{align}
which represents another conceivable technique when utilizing less accurate (with respect to the number of recorded particles and the smallest adjustable class width available) experimental measurement techniques for obtaining a PSD (e.g. laser diffraction analysis). This method shows minor influence on the PSDs at higher resolutions but leads to notable changes compared to the mass-based merging when a larger range of particle fractions is merged. Also for these PSDs, the reaction front speeds are calculated for different equivalence ratios and the results are shown in Fig.~\ref{fig:realPSDs} (right). In general, equal trends can be observed between number-based merging and to mass-based merging. However, the difference to the highly resolved simulation is increased when merging particle fractions by the particle number. Especially lean setups show an increased deviation of \qty{9.9}{\%} (mass based merge: \qty{4.8}{\%}).

In summary, these results illustrate, that already a quite coarse representation of the PSD (in our case: with only 7 particle fractions) can accurately recover the overall reaction front speed over a wide range of equivalence ratios for the iron-air suspension. Nevertheless, considering the original PSD shown in Fig.~\ref{fig:realPSDs_PSDs}, it can be expected that a better resolution is required, when the fraction of smaller particles is larger, especially, for particle fractions with diameters below the smallest adjustable class width. When performing large scale simulations of iron-dust flames, numerical cost (i.~e.~number of considered particle fractions in the PSD and overall transported parcels) have to be weighed against the accuracy for the prediction of the reaction front speed. The results obtained from our investigations on generic and more realistic PSDs can serve as a guiding example for approaching such modeling decisions.

\section{Conclusion}
\label{sec:conclusion}

The primary objective of the present work is to elucidate the influence of polydispersity on reaction front speeds in iron-air suspensions. To this end, three different models (FIRR, FOSK, SOLD) describing the thermochemical oxidation of iron microparticles have been implemented into a flame solver. The code allows arbitrary particle size distributions to be employed and is used for determining the reaction front speed varying the equivalence ratio, PSD, and particle models in a systematic manner. Based on the parametric study the influence of polydispersity on the reaction front speed is analyzed.

The particle models are first validated using single particle experiments by Ning~et~al.~\cite{Ning2021, Ning2022}. The reported temperature profiles are reasonably matched and
all three investigated models show a good agreement with respect to peak temperature and the particles' combustion time. While the single particle experiment is reasonably well predicted, the calculated reaction front speeds are found to vary significantly depending on the used particle model. In agreement with earlier work by Goroshin et al.~\cite{Goroshin1998} the numerical results reported here show that the ignition temperature is an important particle property which significantly influences the reaction front speed. These results suggest that future experimental investigations on the oxidation of micron-sized iron particles should also carefully take the ignition characteristics into account.

Prescribing PSDs with increasing complexity, i.~e.~bidisperse PSDs, generic polydisperse PSDs, and an exemplary PSD measured for a real iron powder sample, it is found that the relative influence of the different particle size fractions on the overall reaction front speed does exhibit a non-trivial relationship to mass or surface of the individual particle fractions. In lean flames, larger particles have an increased influence on the reaction front speed, while in rich flames the smaller particles show a more pronounced effect on the reaction front speed. The equivalence ratio, which marks the transition between these two trends depends on the particle model employed in the simulation. Thus, it is not possible to represent a polydisperse flame by a representative monodisperse equivalent.

An analysis of the reaction front structure for a generic, polydisperse iron-air suspension reveals that the particles ignite sequentially throughout the flame front, depending on their size (i.~e.~their thermal inertia) and the ignition behavior as predicted by the particle model. Smaller particles undergo a thermal runaway more upstream while larger particles are oxidized more downstream which leads to individual combustion environments with differing surrounding gas temperatures and oxygen concentrations for separate particle size fractions. Thus, the position of the thermal runaway, which is primarily dependent on the particle kinetics, also determines the boundary conditions for the particle's burnout. Small particles can burn under excess oxygen in a polydisperse iron flame, even if the overall flame is very fuel rich, and it is observed that the smallest particles undergo intense thermal overshoots, whose intensity decreases with increasing particle size. Large particles, on the other hand, burn under oxygen depleted (richer) conditions compared to the average equivalence ratio of the flame. In general, the combustion environment for individual particles strongly varies within reaction fronts in polydisperse iron-air suspensions and is therefore not necessarily comparable to conditions in common single particle experiments (reduced oxygen concentrations, elevated temperatures). These complex relationships underline that predictive modeling approaches for real iron-dust flames require information beyond single particle experiments, even though such experiments provide very relevant reference data. Model development would significantly benefit from measurements of particle ignition behaviour and reaction front speed measurements for well-characterized iron powders (particle sizes and morphologies, chemical compostion or homogenity), which could serve as calibration targets and validation data. With this, more reliable predictions of polydispersity effects on reaction front speeds in iron-air suspensions could be achieved.

\section*{\hypertarget{appendix}{Appendix: Curvature correction of oxide layer diffusion}}

The original equation by Mi~et~al.~\cite{Mi2022} reads:

\begin{align}
    \difft{m_{\mathrm{FeO}}} & = 
    \frac{A_{\mathrm{FeO}}}{X_{\mathrm{FeO}}}\,\varrho_{\mathrm{FeO}}\,k_{\infty, \mathrm{FeO}}\, \mathrm{exp}\left(-\frac{T_{a,\mathrm{FeO}}}{T_p}\right), 
    \label{eq:SOLD:FeO_plane}
\end{align}

For modeling the full conversion of spherical iron microparticles, Eq.~(\ref{eq:SOLD:FeO_plane}) requires a curvature correction. In their original approach, Mi~et~al.~\cite{Mi2022} focused primarily on the particles' ignition behavior. Prior to ignition, the oxide layer thickness is much smaller than the particle diameter. Therefore, curvature effects are negligibly small and assuming a planar oxide layer is justified. On the contrary, for the iron dust flames studied in this work, the iron core is fully consumed. In this case, the oxide layer thickness reaches the dimension of the particle diameter and the curvature of the oxide layers must be considered~\cite{Fromhold1988}.

\newcommand{\ccFlux}{Q}
\newcommand{\ccPotential}{\Delta P}
\newcommand{\ccResistance}{R}
\newcommand{\ccConductanceCoeff}{D}
\newcommand{\ccResFormFactor}{R_{g}}

According to Wagner's theory, which forms the basis for Eq.~(\ref{eq:SOLD:FeO_plane}), the production rate $\difftinline{m_{\mathrm{FeO}}}$, scales with a diffusive flux through the solid oxide layer which is inversely proportional to its thickness, i.~e.~it represents a \emph{diffusion resistance}. In this semi-empirical approach, the driving potential and diffusivities are aggregated in the kinetic rate~$k$ described by an Arrhenius term~\cite{Atkinson1985}. Only considering diffusion resistances, we note $\difftinline{m_i} \propto A_p/X_i = 1/\ccResFormFactor$, where $R$ is a resistance term with the unit \unit{\permeter} and the expression here describes a planar geometry. For curved geometries, such as cylinders and spheres, $\ccResFormFactor$~requires modification. Such geometry terms for diffusive transport are standardized for heat transfer calculations, e.~g.~\cite{VDI-Heat-Atlas-2010}, and can be used in analogy according to Tab.~\ref{tab:diffusionGeometries}. A comprehensive derivation for diffusive transport of metal ions in microparticles was given by Fromhold~et~al.~\cite{Fromhold1988}. 

\begin{table}
\centering
\caption{Influence of the geometry on the diffusion resistance.}
\label{tab:diffusionGeometries}
\begin{tabularx}{9cm}{
    | >{\centering\arraybackslash}X
    | >{\centering\arraybackslash}X
    | >{\centering\arraybackslash}X| }
    \hline
    Plane & Cylinder & Sphere\\
    \hline
    \includegraphics[width=2.5cm]{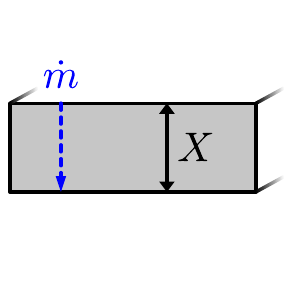}&\includegraphics[width=2.5cm]{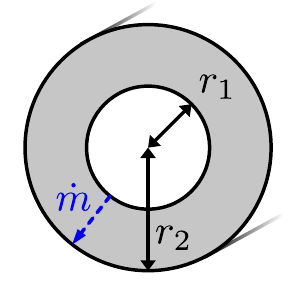}&\includegraphics[width=2.5cm]{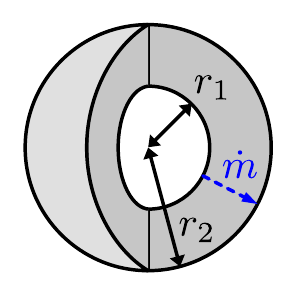}\\
    \hline
    $\ccResFormFactor = \frac{X}{A}$ & 
    $ \ccResFormFactor = \frac{\mathrm{ln}(r_2 / r_1)}{2\pi h}$ & 
    $\ccResFormFactor = \frac{r_2 - r_1}{4 \pi \, r_1 r_2}  
    \phantom{\bigg|}$ \\
    \hline
\end{tabularx}
\end{table}

Equation~(\ref{eq:SOLD:FeO_plane}) is modified by replacing the diffusion resistance term according to Tab.~\ref{tab:diffusionGeometries} for spherical geometry.

\section*{Supplementary material}
The processed, secondary data which is visualized by the plots in this work is accessible in csv format under the DOI:  \href{https://doi.org/10.48328/tudatalib-1081}{10.48328/tudatalib-1081}. For selected simulations, unprocessed raw data is included.

\section*{Acknowledgement}

This work was funded by the Hessian Ministry of Higher Education, Research, Science and the Arts - cluster project Clean Circles. The authors thank Prof. Xiaocheng Mi for fruitful discussions.

\bibliographystyle{ieeetr}
\bibliography{bibliography}

\end{document}